\documentclass{scrartcl}
\usepackage{cite}
\usepackage{float}
\usepackage{sidecap}
\usepackage{subfig}
\usepackage{graphicx}
\usepackage{sidecap}
\usepackage{amsmath}
\usepackage[affil-it]{authblk}

\newcommand{\ket}[1]{| #1 \rangle}

\begin{document}
\title{Landau-Zener transitions in the presence of harmonic noise}

\author[1,2]{Matthias Kraft}
\author[1]{Stephan Burkhardt}
\author[3]{Riccardo Mannella}
\author[1]{Sandro Wimberger}
\affil[1]{\small{Institut f\"ur theoretische Physik, Universit\"at Heidelberg, Philosophenweg 19, 69120 Heidelberg, Germany}}
\affil[2]{\small{Blackett Laboratory, Imperial College London, South Kensington Campus, SW7 2AZ London, UK}}
\affil[3]{\small{Dipartimento di Fisica ‘E. Fermi’, Universit\`a di Pisa, Largo Pontecorvo 3, 56127 Pisa, Italy}}

\maketitle

\begin{abstract}
We study the influence of off-diagonal harmonic noise on transitions
in a Landau-Zener model.
We demonstrate that the harmonic noise can change the transition
probabilities substantially and that its impact depends strongly on
the characteristic frequency of the noise.
In the underdamped regime of the noise process, its effect is compared
with the one of a deterministic sinusoidally oscillating function.
While altering the properties of the noise process allows one to
engineer the transitions probabilities, driving the system with a
deterministic sinusoidal function can result in larger and more
controlled changes of the transition probability.
This may be relevant for realistic implementations of our model with
Bose-Einstein condensates in noise-driven optical lattices.

\end{abstract}

\section{Introduction}\label{sc:Intro}

The Landau-Zener model is one of the simplest, yet non-trivial time-dependent problems
in quantum mechanics. Originally solved in 1932 \cite{Zener, Landau, Majo, Stuckel}, it has
since provided a valuable tool in the study of two state quantum systems \cite{Nikitin,Wernsdorfer}.
In recent years it experienced a renaissance amongst both, theorists
and experimentalists.
On the experimental side, new techniques allowed to test theoretical predictions
with improved precision \cite{Dreisow2009,Longhi, ZenesiniPRL, GhazalPRA10,Regensburger2012}.
On the theoretical side, in-depth studies on finite time effects in Landau-Zener transitions \cite{Vitanov1, Vitanov2} and
on ``noisy'' Landau-Zener transitions \cite{Kayanuma1, Kayanuma2, Pokrovsky} were conducted.

Applications of the Landau-Zener model are versatile; it can provide a tool
to model and manipulate ``qubits'' in circuit QED \cite{Kohler}, to calculate transitions in semiconductor quantum dots \cite{Murgida, Murgida2} or enhance
the experimental control over Bose-Einstein condensates in optical lattices \cite{Ghazal3}. Along those lines,
substantial effort has been made to find new ways to control Landau-Zener transitions by, for example, developing
new experimental protocols \cite{Morsch}. Another approach to control
the transitions, which we will also follow in this letter, is the
introduction of noise into the model \cite{GhazalPRA11}. A different approach to the role of noise in bistable quantum system can be found in \cite{Caldara}.

In this letter, we present a modified version of the Landau-Zener
model containing harmonic noise that directly originates from studying the behavior
of Bose-Einstein condensates in noise-driven optical lattices
(see \cite{NJPPaper} as well as \cite{Lorch}). We demonstrate that in this model, the transition
rates can be effectively controlled by changing the noise properties
and that this control can be even enhanced by replacing the noise
process by a deterministically oscillating phase.

This letter is organized as follows: first we present our modified Landau-Zener model, then we
introduce harmonic noise and discuss its most important properties. In section \ref{sc:NumRes}, we give numerical
results for the diabatic transition probabilities.
We conclude with further perspectives on experimental realizations in
section \ref{sc:Concl}.

\section{``Noisy'' Landau-Zener model and Harmonic Noise}
\subsection{Landau-Zener model with off-diagonal noise}\label{sc:NoisyLZ}
In this section, we will briefly introduce our modified ``noisy'' Landau-Zener model.
The Hamiltonian of our model, motivated by studies of BECs in optical
lattices, reads
\begin{equation}
 \hat{H}_{N, LZ} =\left(
         \begin{array}{cc}
  		  - \frac{F_0}{2} t &  \frac{V_0}{2} (1+e^{i \phi})\\
  					\frac{V_0}{2} (1+e^{-i \phi}) &  \frac{F_0}{2} t \\
         \end{array}\right) \label{HLZNoise} .
\end{equation}
Here, the main difference to the standard Landau-Zener model \cite{Zener, Landau, Majo, Stuckel} is the appearance of a complex phase
on the off-diagonal. As $\phi$ represents our noise variable (more details in section \ref{sc:HarNoise}), this is also the term responsible
for the stochastic nature of our model. The main difference to commonly
studied ``noisy'' Landau-Zener models \cite{Kayanuma1, Kayanuma2} is
that the coupling term on the off-diagonal is composed of a constant,
as well as a noise-driven term. Physically, this model can be used to
analyze Bose-Einstein condensates in tilted optical lattices. In this
case, each of the two terms on the off-diagonal represents a
sinusoidal optical lattice.
While the constant term stands for a static lattice, the complex phase
term stands for a second optical lattice of the same wavelength
\cite{NJPPaper}, shifted by a phase $\phi$.

The states $\ket{0}_{d} = \left(1,0\right)^T$ and
$\ket{1}_{d} = \left(0,1\right)^T$ are called the \emph{diabatic} basis of
the system (eigenstates of $\eqref{HLZNoise}$ with $V_0=0$), while the time-dependent energy eigenstates $\ket{0}_{a}(t)$,$\ket{1}_{a}(t)$ are called the \emph{adiabatic}
basis of the system \cite{Vitanov2}.  The diabatic and the adiabatic eigenstates
are identical in the limit $t=\pm \infty$. However, around $t=0$ the adiabatic eigenstates go through
an avoided crossing, whereas the diabatic eigenstates cross each other. Thus the correspondence
between the two bases ``switches'' such that $\ket{1}_{a}(t=-\infty) = \ket{1}_{d}$, whereas $\ket{1}_{a}(t=+\infty) = \ket{0}_{d}$.
The instantaneous stochastic eigenvalues (adiabatic states) of this model can be found by dia\-gonalizing the Hamiltonian \eqref{HLZNoise}; we then
define an effective band gap\footnote{The band gap is defined as the minimal difference
between the two instantaneous eigenvalues of Hamiltonian \eqref{HLZNoise} and thus a function of the stochastic phase
$\phi(t)$.} at $t=0$ by averaging over the stochastic phase $\phi$ \cite{NJPPaper}. This yields,
\begin{equation}
\mbox{\footnotesize $\Delta$} E_{\rm eff}=V_0 \langle \sqrt{2(\cos(\phi(0))+1)}\rangle \label{EffBG} ,
\end{equation}
where $\langle \dots\rangle$ denotes an average over the equilibrium distribution of $\phi$ (see \eqref{eq:1}).

The most interesting property of this system are the diabatic and adiabatic transition
probabilities. The diabatic transition probability ($P_{\rm d, tra}$) is the probability to end up in $\ket{1}_d$ at $t=+\infty$
if the system was initially prepared in the opposite state $\ket{0}_d$ at $t=-\infty$, or vice versa. The adiabatic transition probabilities accordingly refer to changes of the adiabatic state.
Since the diabatic and adiabatic eigenstates are identical at $t=-\infty$ but are
``swapped'' at $t=+\infty$, the adiabatic transition probability is just $P_{\rm a, tra}=1-P_{\rm d, tra}$.
These transition probabilities can be computed exactly for the original
Landau-Zener problem ($\phi(t)=0$ in equation \eqref{HLZNoise})
\cite{Zener, Landau, Majo, Stuckel}. The diabatic transition
probability reads $P_{\rm d, tra}=1-\exp (-\frac{\pi V_0^2}{2 F_0})$, with $V_0$ being the
band gap at $t=0$.
Using the bandgap given in \eqref{EffBG}, the
formula can also give an estimate of the diabatic transition
probability for the Hamiltonian \eqref{HLZNoise} by replacing $V_0^2$
with $\mbox{\footnotesize $\Delta$} E_{\rm eff}^2$, i.e. $P_{\rm
  est}=1-\exp (-\frac{\pi \mbox{\tiny $\Delta$} E_{\rm eff}^2}{2
  F_0})$.

\subsection{Harmonic Noise}\label{sc:HarNoise}
Previous studies of noise-driven Landau-Zener problems were mainly
concerned with the influence of white \cite{Kayanuma1} or
exponentially correlated noise \cite{Kayanuma2,Pokrovsky}. To better
understand the interplay between noise and the timescales of the
system itself, in this letter the effect of a noise process with a
characteristic frequency is investigated. Here, we will give a short
overview over this so-called \emph{harmonic noise process}.

The Langevin equation defining harmonic noise can be
written as \cite{Geier,Dykman1993}
\begin{align}
  \label{eq:2}
  \begin{split}
    \dot \phi =& \nu \\
    \dot \nu =& -2 \Gamma \nu - \omega_0 ^2 \phi + \sqrt{4\Gamma T}
    \xi(t).
  \end{split}
\end{align}
From these equations, we can see that the harmonic noise process is
nothing but a damped harmonic oscillator driven by a white noise
process $\xi(t)$. The properties of this white noise process are
determined by $\langle \xi(t) \rangle = 0$ and $\langle \xi(t) \xi(t')
\rangle = \delta (t-t')$. The equilibrium distributions of the two
noise variables $\phi$ and $\nu$ are independent and of a gaussian
shape with \cite{Geier}
\begin{align}
  \label{eq:1}
  \langle \phi(t) \rangle &= 0 & \langle \phi(t)^2 \rangle &=
  \frac{T}{\omega_0^2}\\
  \langle \nu(t) \rangle &= 0 & \langle \nu(t)^2 \rangle &= T.
\end{align}
At first it might seem counterintuitive that these distributions do
not depend on the damping coefficient $\Gamma$.
The reason for this is the form of the noise term in \eqref{eq:2},
which is scaled with $\sqrt{\Gamma}$ and thus compensates the
damping.

Many properties of the harmonic noise process can be understood
through the analogy to the damped harmonic oscillator. Just as its
non-stochastic counterpart, the behavior of the harmonic noise can be
separated into two different regimes: the oscillating (under-damped)
and non-oscillating (over-damped) one. If $\omega_0 ^2 > 2\Gamma^2$,
the process shows oscillations, while for the opposite case, its
behavior is closer to an exponentially correlated noise process.

\begin{SCfigure}
  \centering
\begingroup
  \makeatletter
  \providecommand\color[2][]{%
    \GenericError{(gnuplot) \space\space\space\@spaces}{%
      Package color not loaded in conjunction with
      terminal option `colourtext'%
    }{See the gnuplot documentation for explanation.%
    }{Either use 'blacktext' in gnuplot or load the package
      color.sty in LaTeX.}%
    \renewcommand\color[2][]{}%
  }%
  \providecommand\includegraphics[2][]{%
    \GenericError{(gnuplot) \space\space\space\@spaces}{%
      Package graphicx or graphics not loaded%
    }{See the gnuplot documentation for explanation.%
    }{The gnuplot epslatex terminal needs graphicx.sty or graphics.sty.}%
    \renewcommand\includegraphics[2][]{}%
  }%
  \providecommand\rotatebox[2]{#2}%
  \@ifundefined{ifGPcolor}{%
    \newif\ifGPcolor
    \GPcolortrue
  }{}%
  \@ifundefined{ifGPblacktext}{%
    \newif\ifGPblacktext
    \GPblacktexttrue
  }{}%
  \let\gplgaddtomacro\g@addto@macro
  \gdef\gplbacktext{}%
  \gdef\gplfronttext{}%
  \makeatother
  \ifGPblacktext
    \def\colorrgb#1{}%
    \def\colorgray#1{}%
  \else
    \ifGPcolor
      \def\colorrgb#1{\color[rgb]{#1}}%
      \def\colorgray#1{\color[gray]{#1}}%
      \expandafter\def\csname LTw\endcsname{\color{white}}%
      \expandafter\def\csname LTb\endcsname{\color{black}}%
      \expandafter\def\csname LTa\endcsname{\color{black}}%
      \expandafter\def\csname LT0\endcsname{\color[rgb]{1,0,0}}%
      \expandafter\def\csname LT1\endcsname{\color[rgb]{0,1,0}}%
      \expandafter\def\csname LT2\endcsname{\color[rgb]{0,0,1}}%
      \expandafter\def\csname LT3\endcsname{\color[rgb]{1,0,1}}%
      \expandafter\def\csname LT4\endcsname{\color[rgb]{0,1,1}}%
      \expandafter\def\csname LT5\endcsname{\color[rgb]{1,1,0}}%
      \expandafter\def\csname LT6\endcsname{\color[rgb]{0,0,0}}%
      \expandafter\def\csname LT7\endcsname{\color[rgb]{1,0.3,0}}%
      \expandafter\def\csname LT8\endcsname{\color[rgb]{0.5,0.5,0.5}}%
    \else
      \def\colorrgb#1{\color{black}}%
      \def\colorgray#1{\color[gray]{#1}}%
      \expandafter\def\csname LTw\endcsname{\color{white}}%
      \expandafter\def\csname LTb\endcsname{\color{black}}%
      \expandafter\def\csname LTa\endcsname{\color{black}}%
      \expandafter\def\csname LT0\endcsname{\color{black}}%
      \expandafter\def\csname LT1\endcsname{\color{black}}%
      \expandafter\def\csname LT2\endcsname{\color{black}}%
      \expandafter\def\csname LT3\endcsname{\color{black}}%
      \expandafter\def\csname LT4\endcsname{\color{black}}%
      \expandafter\def\csname LT5\endcsname{\color{black}}%
      \expandafter\def\csname LT6\endcsname{\color{black}}%
      \expandafter\def\csname LT7\endcsname{\color{black}}%
      \expandafter\def\csname LT8\endcsname{\color{black}}%
    \fi
  \fi
  \setlength{\unitlength}{0.0500bp}%
  \begin{picture}(3968.00,2834.00)%
    \gplgaddtomacro\gplbacktext{%
      \csname LTb\endcsname%
      \put(1342,704){\makebox(0,0)[r]{\strut{} 1e-06}}%
      \put(1342,1450){\makebox(0,0)[r]{\strut{} 0.0001}}%
      \put(1342,2196){\makebox(0,0)[r]{\strut{} 0.01}}%
      \put(1474,484){\makebox(0,0){\strut{}0}}%
      \put(2166,484){\makebox(0,0){\strut{}$\omega _1$}}%
      \put(2872,484){\makebox(0,0){\strut{}10}}%
      \put(176,1636){\rotatebox{-270}{\makebox(0,0){\strut{}$S_{\Phi}(\omega)$}}}%
      \put(2522,154){\makebox(0,0){\strut{}$\omega$}}%
    }%
    \gplgaddtomacro\gplfronttext{%
    }%
    \gplbacktext
    \put(0,0){\includegraphics{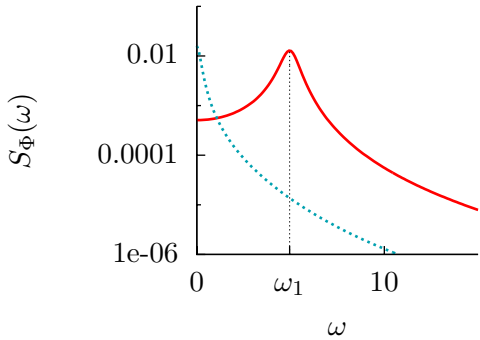}}%
    \gplfronttext
  \end{picture}%
\endgroup
  \caption{Power spectrum of a harmonic noise process in the
    under-damped (red solid line) and over-damped case (dotted blue
    line). Noise parameters are: $T=1.0,\Gamma=0.5,\omega_0 =
    5.0$ (under-damped case) and $T=0.01,\Gamma=2.5,\omega_0 =
    1.0$ (over-damped case).}
\label{fig:harmspectrum}
\end{SCfigure}

The difference between these two regimes becomes evident when looking
at the power spectrum of the noise variable $\phi(t)$. This power
spectrum is given by
\begin{equation}
  \label{eq:3}
  S_\phi (\omega) = \frac{2\Gamma T}{\pi (4\Gamma^2 \omega^2 +
    (\omega^2 - \omega_0 ^2)^2)},
\end{equation}
and can be seen in figure \ref{fig:harmspectrum}. For the oscillating
case, we can see that this response function does
have a clear maximum at $\omega_1 = \sqrt{\omega_0 ^2 - 2 \Gamma^2}$,
while for the non-oscillating case, the maximum is assumed at
$\omega=0$. For small damping coefficients, the width of the maximum
is proportional to $2 \Gamma$. This power spectrum is very similar  to
the response function of a damped harmonic oscillator driven with the
frequency $\omega$.

Without damping ($\Gamma = 0$), \eqref{eq:2} turns into the
differential equation of a harmonic oscillator and the power spectrum
\eqref{eq:3} correspondingly turns into a delta distribution
$\delta(\omega-\omega_0)$. If the damping time $1/\Gamma$ is very
small compared to the timescale of the system, the noise process is
approximated by a simple harmonic oscillation. In order to conserve
the total power (integral over the power spectrum) of the process,
this approximation should take the following form \cite{Matthias}:
\begin{equation}
\phi(t)_{det} = 2\sqrt{\frac{T}{\omega_0^2}} \sin(\omega_0 t + \phi_0)=2\sqrt{\rm Var(\phi)} \sin(\omega_0 t + \phi_0).
\label{eq:4}
\end{equation}
\section{Numerical Results}\label{sc:NumRes}
In this section, we present numerical results for the diabatic transition probability in dependence of the characteristic frequency of 
harmonic noise. The results have been obtained by numerically integrating the time-dependent Schr\"odinger equation corresponding to the
Hamiltonian \eqref{HLZNoise}; the initial state was chosen to be the diabatic ground state of the model. The integration time was long
compared to the transition time \cite{Vitanov1, Vitanov2} such that finite time effects were minimal. We averaged the diabatic transition probability
over the last ten percent of the integration time to estimate its asymptotic value at $t=\infty$. For details see \cite{Matthias}. In the
following ``transition probability'' will always refer to the diabatic transition probability.
\begin{figure}[!h]
\begingroup
  \makeatletter
  \providecommand\color[2][]{%
    \GenericError{(gnuplot) \space\space\space\@spaces}{%
      Package color not loaded in conjunction with
      terminal option `colourtext'%
    }{See the gnuplot documentation for explanation.%
    }{Either use 'blacktext' in gnuplot or load the package
      color.sty in LaTeX.}%
    \renewcommand\color[2][]{}%
  }%
  \providecommand\includegraphics[2][]{%
    \GenericError{(gnuplot) \space\space\space\@spaces}{%
      Package graphicx or graphics not loaded%
    }{See the gnuplot documentation for explanation.%
    }{The gnuplot epslatex terminal needs graphicx.sty or graphics.sty.}%
    \renewcommand\includegraphics[2][]{}%
  }%
  \providecommand\rotatebox[2]{#2}%
  \@ifundefined{ifGPcolor}{%
    \newif\ifGPcolor
    \GPcolortrue
  }{}%
  \@ifundefined{ifGPblacktext}{%
    \newif\ifGPblacktext
    \GPblacktexttrue
  }{}%
  \let\gplgaddtomacro\g@addto@macro
  \gdef\gplbacktext{}%
  \gdef\gplfronttext{}%
  \makeatother
  \ifGPblacktext
    \def\colorrgb#1{}%
    \def\colorgray#1{}%
  \else
    \ifGPcolor
      \def\colorrgb#1{\color[rgb]{#1}}%
      \def\colorgray#1{\color[gray]{#1}}%
      \expandafter\def\csname LTw\endcsname{\color{white}}%
      \expandafter\def\csname LTb\endcsname{\color{black}}%
      \expandafter\def\csname LTa\endcsname{\color{black}}%
      \expandafter\def\csname LT0\endcsname{\color[rgb]{1,0,0}}%
      \expandafter\def\csname LT1\endcsname{\color[rgb]{0,1,0}}%
      \expandafter\def\csname LT2\endcsname{\color[rgb]{0,0,1}}%
      \expandafter\def\csname LT3\endcsname{\color[rgb]{1,0,1}}%
      \expandafter\def\csname LT4\endcsname{\color[rgb]{0,1,1}}%
      \expandafter\def\csname LT5\endcsname{\color[rgb]{1,1,0}}%
      \expandafter\def\csname LT6\endcsname{\color[rgb]{0,0,0}}%
      \expandafter\def\csname LT7\endcsname{\color[rgb]{1,0.3,0}}%
      \expandafter\def\csname LT8\endcsname{\color[rgb]{0.5,0.5,0.5}}%
    \else
      \def\colorrgb#1{\color{black}}%
      \def\colorgray#1{\color[gray]{#1}}%
      \expandafter\def\csname LTw\endcsname{\color{white}}%
      \expandafter\def\csname LTb\endcsname{\color{black}}%
      \expandafter\def\csname LTa\endcsname{\color{black}}%
      \expandafter\def\csname LT0\endcsname{\color{black}}%
      \expandafter\def\csname LT1\endcsname{\color{black}}%
      \expandafter\def\csname LT2\endcsname{\color{black}}%
      \expandafter\def\csname LT3\endcsname{\color{black}}%
      \expandafter\def\csname LT4\endcsname{\color{black}}%
      \expandafter\def\csname LT5\endcsname{\color{black}}%
      \expandafter\def\csname LT6\endcsname{\color{black}}%
      \expandafter\def\csname LT7\endcsname{\color{black}}%
      \expandafter\def\csname LT8\endcsname{\color{black}}%
    \fi
  \fi
  \setlength{\unitlength}{0.0500bp}%
  \begin{picture}(7142.00,4414.00)%
    \gplgaddtomacro\gplbacktext{%
      \csname LTb\endcsname%
      \put(946,2711){\makebox(0,0)[r]{\strut{} 0.4}}%
      \put(946,3190){\makebox(0,0)[r]{\strut{} 0.6}}%
      \put(946,3670){\makebox(0,0)[r]{\strut{} 0.8}}%
      \put(946,4149){\makebox(0,0)[r]{\strut{} 1}}%
      \put(1078,2251){\makebox(0,0){\strut{}}}%
      \put(2495,2251){\makebox(0,0){\strut{}}}%
      \put(3912,2251){\makebox(0,0){\strut{}}}%
      \put(5328,2251){\makebox(0,0){\strut{}}}%
      \put(6745,2251){\makebox(0,0){\strut{}}}%
      \put(176,3310){\rotatebox{-270}{\makebox(0,0){\strut{}$P_{\mathrm{d, tra}}(t=\infty)$}}}%
    }%
    \gplgaddtomacro\gplfronttext{%
    }%
    \gplgaddtomacro\gplbacktext{%
      \csname LTb\endcsname%
      \put(946,937){\makebox(0,0)[r]{\strut{} 0.4}}%
      \put(946,1404){\makebox(0,0)[r]{\strut{} 0.6}}%
      \put(946,1871){\makebox(0,0)[r]{\strut{} 0.8}}%
      \put(946,2338){\makebox(0,0)[r]{\strut{} 1}}%
      \put(1078,484){\makebox(0,0){\strut{} 0}}%
      \put(2495,484){\makebox(0,0){\strut{} 1.5}}%
      \put(3912,484){\makebox(0,0){\strut{} 3}}%
      \put(5328,484){\makebox(0,0){\strut{} 4.5}}%
      \put(6745,484){\makebox(0,0){\strut{} 6}}%
      \put(176,1521){\rotatebox{-270}{\makebox(0,0){\strut{}$P_{\mathrm{d, tra}}(t=\infty)$}}}%
      \put(3911,154){\makebox(0,0){\strut{}$\omega_0$}}%
    }%
    \gplgaddtomacro\gplfronttext{%
    }%
    \gplbacktext
    \put(0,0){\includegraphics{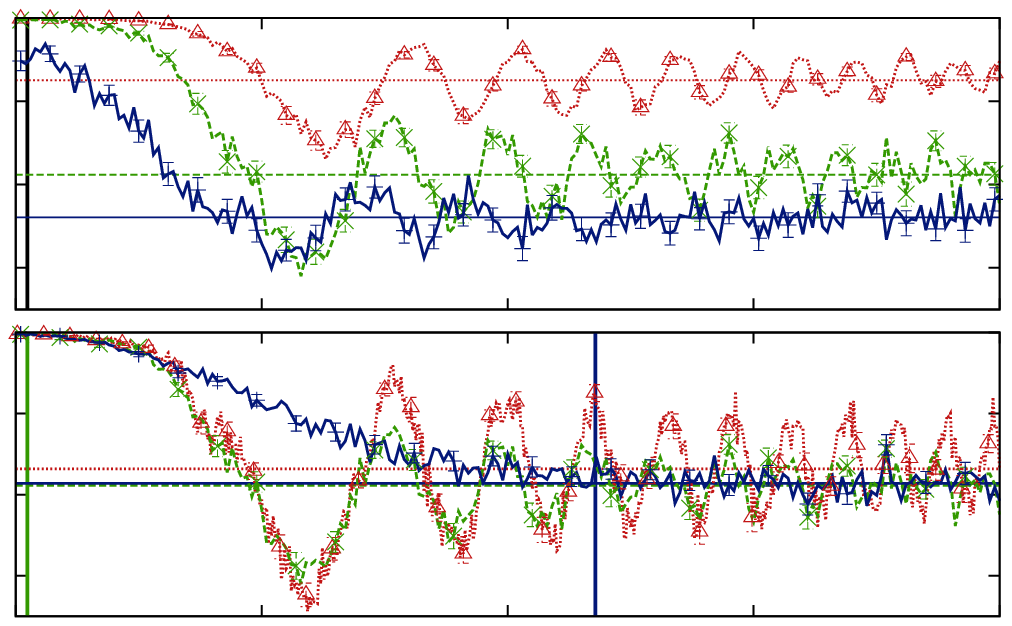}}%
    \gplfronttext
  \end{picture}%
\endgroup
  \caption{Diabatic transition probability versus the dimensionless
    noise frequency $\omega_0$. The top figure shows results for $\rm
    Var(\phi)=0.1$ (red dotted, triangles),
  $\rm Var(\phi)=0.5$ (green dashed, crosses) and $\rm Var(\phi)=2$
  (blue solid, lines). The other parameters are $F_0=0.5, \Gamma=0.05$, and $V_0=0.75$. The vertical solid black line at $\omega_0=0.07$ marks 
  the point after which the noise shows oscillatory behavior. The horizontal lines give the saturation value around which the transition probability
  oscillates. \\
  The bottom figure shows results for $\Gamma=0.0$ (red dotted,
  triangles), $\Gamma=0.05$ (green dashed, crosses) and $\Gamma=2.5$
  (blue solid, lines). 
  The other parameters are $F_0=0.5,\rm Var(\phi)=0.5$, and $V_0=0.75$. The vertical lines (solid green at $\omega_0=0.07$ for $\Gamma=0.05$ and solid blue
  at $\omega_0=3.53$ for $\Gamma=2.5$)
  mark the points after which the noise shows oscillatory behavior. The horizontal lines again give the saturation value of the
  transition probability. \\
  The error bars are only shown for every 8th point and give the error of the expectation value.
  }\label{fig:Num1}
\end{figure}

Figure \ref{fig:Num1} shows the transition probability versus the dimensionless noise frequency $\omega_0$ for different variances (top) and damping
coefficients $\Gamma$ (bottom).
The transition probability strongly depends on the noise
frequency. Comparison with the transition probability $P_{\rm
  est}\approx0.989-0.999$ (for all graphs) estimated using the
effective band gap from \eqref{EffBG}, reveals that harmonic noise
strongly reduces the diabatic transition probability, i.e. it
increases the adiabatic transition probability ($P_{\rm a,tra}=1-P_{\rm d,
  tra}$) to the upper band.
This opens up the possibility to influence the transition probability
by changing the noise parameters (see also \cite{GhazalPRA11}).

Taking a more detailed look at figure \ref{fig:Num1}, we see that for all six data sets the influence of the noise is small for small frequencies, but the transition probabilities strongly decrease between
$\omega_0\approx0.5-1.5$ leading to clear local minima (apart from the data with $\Gamma=2.5$). The positions of those minima depend linearly on the effective band gap; this effect is studied in
detail in \cite{NJPPaper}. 

Disregarding the data for $\Gamma=2.5$, the transition probability oscillates around a saturation value for increasing $\omega_0$. 
Figure \ref{fig:Num1} (top) shows that this saturation value depends on the variance of the noise process, i.e. on the noise strength. For small
variances the noise influence is relatively small; increasing the variance enhances the
noise influence, until the saturation value drops to $\approx1/2$ for $\rm Var(\phi)=2$. Strong noise thus leads to an incoherent superposition of the
two energy states, a fact already discovered by Kayanuma \cite{Kayanuma1, Kayanuma2} for similar LZ models. This supports the intuitive notion
that strong noise leads to a complete mixing of the energy levels where the band gap is minimal.

Figure \ref{fig:Num1} (bottom) shows that changing the damping coefficient $\Gamma$ does not change the saturation value as much. However, it can be seen that changing the damping coefficient
controls the sharpness and height of the oscillations around the saturation value (they are completely damped out for $\Gamma=2.5$). 

Analyzing the two extremes, $\Gamma=0$ ($\phi$ describes sinusoidal motion and we average over amplitude and phase according to \eqref{eq:1}) and $\Gamma=2.5$,
we can see that the transition probability for $\Gamma=0$ still shows the same characteristic decrease and oscillations. The transition probability
for $\Gamma=2.5$ does not show those oscillations and the initial decrease happens at a slower rate. Nevertheless, the saturation value is almost
the same as for $\Gamma=0.05$.
We can thus conclude that the existence of this saturation value is a
general feature of harmonic noise, independent of the noise parameters
we are working with.
On the other hand, the \emph{rapid} initial decrease in the transition
probability and the oscillations around a saturation value are due to
the sinusoidal nature of harmonic noise.
This leads us to the next figure, in which we compare harmonic noise
with a deterministically oscillating phase.
\begin{figure}[t]
 \centering
\begingroup
  \makeatletter
  \providecommand\color[2][]{%
    \GenericError{(gnuplot) \space\space\space\@spaces}{%
      Package color not loaded in conjunction with
      terminal option `colourtext'%
    }{See the gnuplot documentation for explanation.%
    }{Either use 'blacktext' in gnuplot or load the package
      color.sty in LaTeX.}%
    \renewcommand\color[2][]{}%
  }%
  \providecommand\includegraphics[2][]{%
    \GenericError{(gnuplot) \space\space\space\@spaces}{%
      Package graphicx or graphics not loaded%
    }{See the gnuplot documentation for explanation.%
    }{The gnuplot epslatex terminal needs graphicx.sty or graphics.sty.}%
    \renewcommand\includegraphics[2][]{}%
  }%
  \providecommand\rotatebox[2]{#2}%
  \@ifundefined{ifGPcolor}{%
    \newif\ifGPcolor
    \GPcolortrue
  }{}%
  \@ifundefined{ifGPblacktext}{%
    \newif\ifGPblacktext
    \GPblacktexttrue
  }{}%
  \let\gplgaddtomacro\g@addto@macro
  \gdef\gplbacktext{}%
  \gdef\gplfronttext{}%
  \makeatother
  \ifGPblacktext
    \def\colorrgb#1{}%
    \def\colorgray#1{}%
  \else
    \ifGPcolor
      \def\colorrgb#1{\color[rgb]{#1}}%
      \def\colorgray#1{\color[gray]{#1}}%
      \expandafter\def\csname LTw\endcsname{\color{white}}%
      \expandafter\def\csname LTb\endcsname{\color{black}}%
      \expandafter\def\csname LTa\endcsname{\color{black}}%
      \expandafter\def\csname LT0\endcsname{\color[rgb]{1,0,0}}%
      \expandafter\def\csname LT1\endcsname{\color[rgb]{0,1,0}}%
      \expandafter\def\csname LT2\endcsname{\color[rgb]{0,0,1}}%
      \expandafter\def\csname LT3\endcsname{\color[rgb]{1,0,1}}%
      \expandafter\def\csname LT4\endcsname{\color[rgb]{0,1,1}}%
      \expandafter\def\csname LT5\endcsname{\color[rgb]{1,1,0}}%
      \expandafter\def\csname LT6\endcsname{\color[rgb]{0,0,0}}%
      \expandafter\def\csname LT7\endcsname{\color[rgb]{1,0.3,0}}%
      \expandafter\def\csname LT8\endcsname{\color[rgb]{0.5,0.5,0.5}}%
    \else
      \def\colorrgb#1{\color{black}}%
      \def\colorgray#1{\color[gray]{#1}}%
      \expandafter\def\csname LTw\endcsname{\color{white}}%
      \expandafter\def\csname LTb\endcsname{\color{black}}%
      \expandafter\def\csname LTa\endcsname{\color{black}}%
      \expandafter\def\csname LT0\endcsname{\color{black}}%
      \expandafter\def\csname LT1\endcsname{\color{black}}%
      \expandafter\def\csname LT2\endcsname{\color{black}}%
      \expandafter\def\csname LT3\endcsname{\color{black}}%
      \expandafter\def\csname LT4\endcsname{\color{black}}%
      \expandafter\def\csname LT5\endcsname{\color{black}}%
      \expandafter\def\csname LT6\endcsname{\color{black}}%
      \expandafter\def\csname LT7\endcsname{\color{black}}%
      \expandafter\def\csname LT8\endcsname{\color{black}}%
    \fi
  \fi
  \setlength{\unitlength}{0.0500bp}%
  \begin{picture}(7142.00,3428.00)%
    \gplgaddtomacro\gplbacktext{%
      \csname LTb\endcsname%
      \put(946,704){\makebox(0,0)[r]{\strut{} 0}}%
      \put(946,1196){\makebox(0,0)[r]{\strut{} 0.2}}%
      \put(946,1688){\makebox(0,0)[r]{\strut{} 0.4}}%
      \put(946,2179){\makebox(0,0)[r]{\strut{} 0.6}}%
      \put(946,2671){\makebox(0,0)[r]{\strut{} 0.8}}%
      \put(946,3163){\makebox(0,0)[r]{\strut{} 1}}%
      \put(1078,484){\makebox(0,0){\strut{} 0}}%
      \put(2495,484){\makebox(0,0){\strut{} 1.5}}%
      \put(3912,484){\makebox(0,0){\strut{} 3}}%
      \put(5328,484){\makebox(0,0){\strut{} 4.5}}%
      \put(6745,484){\makebox(0,0){\strut{} 6}}%
      \put(176,1933){\rotatebox{-270}{\makebox(0,0){\strut{}$P_{\mathrm{d, tra}}(t=\infty)$}}}%
      \put(3911,154){\makebox(0,0){\strut{}$\omega_0$}}%
    }%
    \gplgaddtomacro\gplfronttext{%
    }%
    \gplbacktext
    \put(0,0){\includegraphics{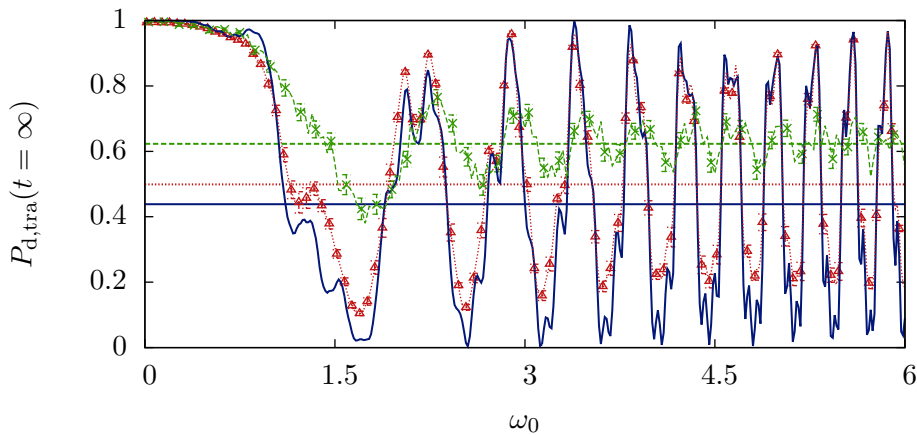}}%
    \gplfronttext
  \end{picture}%
\endgroup
  \caption{Diabatic transition probability versus the dimensionless noise frequency $\omega_0$. We show data for the ``noisy'' LZ-model with harmonic noise
  ($\Gamma=0.05, \rm Var(\phi)=0.5$)
  (green dashed line, crosses), with a deterministic phase
  $\phi=\sqrt{2}\sin(\omega_0 t + \phi_0)$ averaged over $\phi_0$(red
  dotted line, triangles) and a single realization of the
  deterministic phase noise with $\phi_0=0.0$ (blue solid line). The horizontal lines give the
  saturation value (dashed green line for the noisy LZ model with harmonic noise, red
dotted line for the deterministic phase and blue solid line for the
single realization of the deterministic phase). For all three, $V_0=0.75$ and $F_0=0.5$. $\phi_0$ has been chosen from
  a flat distribution between $0$ and $2\pi$. We averaged over 100 realizations. Error bars as in fig.\ref{fig:Num1}.}\label{fig:Num2}
\end{figure}

Figure \ref{fig:Num2} shows the transition probability versus the dimensionless frequency $\omega_0$ for a harmonic noise process and a
deterministically oscillating phase as given in Eq. \eqref{eq:4} (we give both an average over $\phi_0$ and a single realization with $\phi_0=0$).
It can be seen that there is a good qualitative agreement between the data obtained with harmonic noise and with a deterministic phase. 
The LZ model with a deterministic phase nicely reproduces the initial decrease in the transition probability and the frequency of the
oscillations around the saturation value. However, the oscillations
are much more prominent for the deterministic phase and the positions of the first few minima/maxima seem slightly shifted to lower values of the frequency.

A possible explanation for this shift and why it vanishes at high frequencies could be 
that the noise does not behave oscillatory at those minima yet and that our model thus breaks down. However, the noise already shows
oscillatory behaviour for $\omega_0>\sqrt{2}\Gamma=0.07$, so it will clearly be highly oscillatory at frequencies
where the minima/maxima occur (first minimum is at $\omega_0\approx1.6$). One will reach the same conclusion
by looking at the power spectrum of the harmonic noise \eqref{eq:3}. For the parameters in figure \ref{fig:Num2} and
with $\omega_0=1.6$, the power spectrum is very narrowly peaked at $\omega_1 = \sqrt{\omega_0^2 -2 \Gamma^2}=1.5992$ and almost completely symmetric,
thus comparing very well with the ``delta peaked power'' spectrum of the deterministic phase. This can, therefore, not be
the reason for the discrepancies.

The real shortcoming of our model is a different one, namely that the amplitude of harmonic noise varies, whereas
the one of the deterministic phase stays constant. 
Preliminary numerical calculations have shown that different amplitudes of the deterministic phase
lead to slightly different transition probabilities, even though the overall oscillatory
behaviour and the approximate position of the minima/maxima stays intact.
Harmonic noise should, in the highly oscillatory regime, thus be more accurately described by a harmonic oscillator with a randomly distributed amplitude that follows a gaussian distribution.
However, for the sake of simplicity of our model, we refrained from introducing a stochastic amplitude.

Amplitude and sharpness of the oscillations render the deterministic phase model even more suited to control the transition probability in the ``noisy'' LZ model
than harmonic noise. Focussing on the transition probability around $\omega_0\approx 3$, we see that a slight change in the frequency results in a huge amplification (reduction)
of the transition probability. In numbers, the transition probability rises from $0.153$ at $\omega_0=3.12$ to $0.958$ at $\omega_0=3.39$ for the deterministic
phase model averaged over the random phase $\phi_0$. This effect will be greatly enhanced
if one is furthermore able to control the random phase $\phi_0$. In this case, the transition probability rises from $0.005$ at $\omega_0=3.12$ to $0.997$ at $\omega_0=3.39$
(about 200 times the value at $\omega_0=3.12$). Even though the difference in the transition probability between subsequent minima and maxima is not always
as prominent, it is still obvious that one can adjust the transition probability over
orders of magnitude by only slightly changing the frequency of the deterministic phase.
It should be noted that we obtained these graphs without any
optimization to find the ``best'' parameters that maximize the impact
of the harmonic noise or the deterministic phase.

\section{Conclusions}\label{sc:Concl}

In this letter, we have shown that the here proposed Landau-Zener type
model allows for greater control of the transition probabilities
between the two energy states when compared to the standard
Landau-Zener model \cite{Zener,Majo,Stuckel,Landau}.
Driving the system with harmonic noise, we were able to change the
transition probabilities while keeping the coupling term $V_0$ as well
as the driving $F_0$ constant and only changing the properties of the
noise.
Compared to previous studies of noise-driven Landau-Zener problems
\cite{Kayanuma1, Kayanuma2, Pokrovsky}, we were able to observe a wide
range of behavior, as seen in figure \ref{fig:Num1}.
We demonstrated that the effects most relevant for controlling the
transitions, such as the dependence of the transition probability on noise
frequency, are due to the sinusoidal nature of the harmonic noise
process (see figure \ref{fig:Num2}).

While the system under the influence of harmonic noise allows to
change the transition probability by tuning the properties of the
noise process, this can be done even more efficiently by using a
deterministically oscillating phase (see Eq. \eqref{eq:4}).
Our numerical simulations revealed that using this deterministic
phase, even small changes in the frequency cause the diabatic transition
probability to change over orders of magnitude.
This, of course, provides the chance 
to develop new experimental protocols (see also \cite{Morsch}) that allow for an excellent control over the system parameters. 
An ideal playground to implement such protocols is given by optical
realizations \cite{Dreisow2009,Regensburger2012} or Bose-Einstein condensates in
optical lattices \cite{ZenesiniPRL,GhazalPRA10,Ghazal3,Morsch}.

\section*{Acknowledgments}
We are grateful for the support of the Excellence Initiative through
the Heidelberg Graduate School of Fundamental Physics (Grant No. GSC
129/1), the DFG FOR760, the Heidelberg Center for Quantum Dynamics and
the Alliance Program of the Helmholtz Association (HA216/EMMI).

\end{document}